\documentclass[nofootinbib]{revtex4}
\usepackage{graphicx}
\usepackage{latexsym}

\def\be{\begin{equation}}
\def\ee{\end{equation}}
\def\bea{\begin{eqnarray}}
\def\eea{\end{eqnarray}}
\newcommand{\DE}{{\mathrm{de}}}

\begin{document}

\title{Impacts on Cosmological Constraints from Degeneracies}

\author{Hong Li${}^{a,b}$}
\email{hongli@ihep.ac.cn}
\author{Jun-Qing Xia${}^{c,a}$}
\email{xia@sissa.it}

\affiliation{${}^a$Key Laboratory of Particle Astrophysics,
Institute of High Energy Physics, Chinese Academy of Science,
P.O.Box 918-3, Beijing 100049, P.R.China}

\affiliation{${}^b$National Astronomical
Observatories, Chinese Academy of Sciences, Beijing 100012, P.~R.~China}

\affiliation{${}^c$Scuola Internazionale Superiore di Studi Avanzati, Via Bonomea 265, I-34136 Trieste, Italy}

%\date{\today}

\begin{abstract}
In this paper, we study the degeneracies among several cosmological parameters in detail and discuss their impacts on the determinations of these parameters from the current and future observations. By combining the latest data sets, including type-Ia supernovae ``Union2.1'' compilation, WMAP seven-year data and the baryon acoustic oscillations from the SDSS Data Release Seven, we perform a global analysis to determine the cosmological parameters, such as the equation of state of dark energy $w$, the curvature of the universe $\Omega_k$, the total neutrino mass $\sum{m_\nu}$, and the parameters associated with the power spectrum of primordial fluctuations ($n_s$, $\alpha_s$ and $r$). We pay particular attention on the degeneracies among these parameters and the influences on their constraints, by with or without including these degeneracies, respectively. We find that $w$ and $\Omega_k$ or $\sum{m_\nu}$ are strongly correlated. Including the degeneracies will significantly weaken the constraints. Furthermore, we study the capabilities of future observations and find these degeneracies can be broken very well. Consequently, the constraints of cosmological parameters can be improved dramatically.
\end{abstract}

%\pacs{98.80.Es; 98.80.Cq}

\maketitle

%Introduction==========================================================

\section{Introduction}

\label{Int}

Measuring cosmological parameters with observational data plays a crucial role in establishing the modern cosmology. In the past few years, with great progresses on astronomical observations, such as type-Ia supernovae (SN) \cite{union2.1}, cosmic microwave background radiation (CMB) \cite{wmap7} and large scale structure (LSS) \cite{lss}, a standard cosmological model is build up. It is commonly believed that our universe has undergone an inflationary period at the beginning and is accelerating expansion at present. Dark energy, the mysterious source driving the present acceleration of our universe, has been studied widely in the literature since its discovery in 1998 \cite{sn1998}. However, the nature of dark energy, encoded in its equation of state (EoS) parameter $w$, remains controversial. For the other acceleration in the very early universe, the mechanics of inflation can naturally explain the flatness, homogeneity and isotropy of our universe. Inflation stretches the primordial density fluctuations and seeds the presently observed CMB and LSS. Understanding the dynamics of inflation and the origin of late-time acceleration are big challenges in physics and astronomy.

With the accumulation of observational data from CMB, LSS and SN observations and the improvements of the data quality,
there are a lot of efforts in the literature on constraining the cosmological parameters from various
observations \cite{wmap7,Xia:2006cr,constraints}, such as the EoS of dark energy models $w$,
the curvature of the universe $\Omega_k$, the total neutrino masses $\sum{m_\nu}$,
and those associated with the running of the spectral index and gravitational waves.
However, using different input theoretical model can affect constraints of cosmological parameters,
since some of cosmological parameters are correlated with each other.
Including or ignoring these degeneracies will give significantly different constraints on cosmological parameters.
For example, the EoS of dark energy $w$ correlates with the matter energy density $\Omega_m$ and the curvature $\Omega_k$ through
the geometrical distance relations. Due to the strong correlation between $w$ and $\Omega_k$,
we obtain different constraints on dark energy EoS $w$ in the framework of flat and non-flat universe,
respectively \cite{Zhao:2006qg}. %Meanwhile, the EoS from different dark energy model also influences the measurements of $\Omega_m$ and $\Omega_k$.
The EoS of dark energy also anti-correlates with the total neutrino mass, namely, the larger neutrino mass can be mimicked by the more negative $w$ for the same distance relations \cite{Xia:2006wd,Li:2008vf,Hannestad:2005gj}. In the framework of dynamical dark energy model, the constraint on $\sum{m_\nu}$ is significantly weakened, when comparing with that obtained from the standard $\Lambda$CDM model. Furthermore, the tensor perturbation mode and the dark energy component are correlated, since they mostly affect the large scale (low multipoles) temperature power spectrum of CMB. Thus, including the tensor fluctuations in the analysis will change the constraint of $w$. Moreover, some other cosmological parameters, such as, the initial conditions, inflationary parameters, can also influence the numerical results through the degeneracies.

At present, we have a lot of cosmological observations, such as several hundred SN samples, CMB temperature and polarization fluctuation, LSS surveys and so forth. It is possible for us to study the determination of cosmological parameters precisely, especially the parameters correlating with $w$. Thus, in this paper, we study the uncertainties of the constraining cosmological parameters in detail, due to the correlations among them. By employing the Markov Chain Monte Carlo method, we combine the latest observational data, such as seven-year WMAP data (WMAP7), baryon acoustic oscillation (BAO) and the ``Union2.1'' compilation SN sample to constrain cosmological parameters in different input cosmological models, such as the non-flat universe, the massive neutrino mass models, and so on. We pay particular attention on the degeneracies among these parameters. Furthermore, we also generate the mock data for the future observations, such as PLANCK \cite{planck} and Euclid \cite{Euclid,Euclid-official} projects, as well as future SN observations \cite{futureSN} to study their capabilities on breaking these degeneracies.

The structure of the paper is as follows: in section \ref{Method} we describe the method and the current and future observational data sets we use; Section \ref{results} contains our main numerical results on the cosmological parameters, and the last section is the summary.

%Method and Current Observations=======================================

\section{Method and Data}
\label{Method}
\subsection{Method}
In order to get constraints in different cosmological models, we employ the modified version of the MCMC package {\tt CosmoMC} \cite{CosmoMC} and perform the global fitting analysis with the current and future observational data. We parameterize the cosmology with the following parameters:
\begin{equation}
\label{parameter} {\bf P} \equiv (\omega_{b}, \omega_{c},
\Omega_k, \Theta_{s}, \tau, w_{0}, w_{a}, f_{\nu}, n_{s}, A_{s},
\alpha_s, r)~,
\end{equation}
where $\omega_{b}\equiv\Omega_{b}h^{2}$ and
$\omega_{c}\equiv\Omega_{c}h^{2}$, in which $\Omega_{b}$ and
$\Omega_{c}$ are the physical baryon and cold dark matter
densities relative to the critical density, $\Omega_k$ is the
spatial curvature and satisfies
$\Omega_k+\Omega_b+\Omega_c+\Omega_{\DE}=1$, $\Theta_{s}$ is the ratio
(multiplied by 100) of the sound horizon to the angular diameter
distance at decoupling, $\tau$ is the optical depth to
re-ionization, $f_{\nu}$ is the dark matter neutrino fraction at present\cite{Lesgourgues:2006nd}, namely,
\begin{equation}
f_{\nu}\equiv\frac{\rho_{\nu}}{\rho_{m}}=\frac{\Sigma
m_{\nu}}{93.14~\mathrm{eV}~\Omega_mh^2}~.
\end{equation}
The primordial scalar power spectrum $\mathcal{P}_{s}(k)$ is parameterized as \cite{Ps}:
\begin{eqnarray}
\ln\mathcal{P}_{s}(k)=\ln
A_s(k_{s0})+(n_s(k_{s0})-1)\ln\left(\frac{k}{k_{s0}}\right)+\frac{\alpha_s}{2}
\left[\ln\left(\frac{k}{k_{s0}}\right)\right]^2
\end{eqnarray}
where $A_s$ is defined as the amplitude of initial power spectrum, $n_s$ measures the spectral index, $\alpha_{s}$ is the running of the scalar spectral index and $r$ is the tensor to scalar ratio of the primordial spectrum. For the pivot scale we set $k_{s0}=0.05\,$Mpc$^{-1}$  which is the default value of {\tt CosmoMC} package. $w_0$ and $w_a$ are the parameters of dark energy, and the EoS is parameterized as
\cite{Linderpara}:
\begin{equation}
\label{EOS} w_\DE(a) = w_{0} + w_{a}(1-a)~,
\end{equation}
(noted as ``CPL''), where $a=1/(1+z)$ is the scale factor and it covers $\Lambda$CDM model by taking $w_0=-1$ and $w_a=0$. As we know, for time evolving dark energy model, it is crucial to include dark energy perturbations in the global fitting strategy to constrain cosmological parameters \cite{WMAP3GF,LewisPert,XiaPert}. In this paper we use the method provided in refs.\cite{XiaPert,ZhaoPert} to treat the dark energy perturbations consistently in the whole parameter space in the numerical calculations.

\subsection{Observational Data}
\subsubsection{Current Data}
In our analysis, we consider the following cosmological probes: i) power spectra of CMB temperature and polarization anisotropies; ii) the baryon acoustic oscillation in the galaxy power spectra; iii) measurement of the current Hubble constant; iv) luminosity distances of type Ia supernovae.

To incorporate the WMAP7 CMB temperature and polarization power spectra, we use the routines for computing the likelihood supplied by the WMAP team \cite{wmaplike}. The WMAP7 polarization data are composed of TE/EE/BB power spectra on large scales ($2 \leq\ell\leq 23$) and TE power spectra on small scales ($24 \leq\ell\leq 800$), while the WMAP7 temperature data includes the CMB anisotropies on scales $2 \leq\ell\leq 1200$.

 Baryon Acoustic Oscillations provides an efficient method for measuring the expansion history by using features in the clustering of galaxies within large scale surveys as a ruler with which to measure the distance-redshift relation. The BAO aries because that the coupling of baryons and photons by Thomson scattering in the early universe allows acoustic oscillations in the plasma at early time. After photons decoupling, the acoustic waves can propagate a certain distance, which becomes a characteristic comoving scale and remains in the distribution of galaxies. It provides a particularly roburst quantity to measure\cite{Peebles:1970ag,Sunyaev:1970eu,Eisenstein:1997ik,Meiksin:1998ra,Eisenstein:2006nj}.  BAO has been detected in the current galaxy redshift survey data from the SDSS DR7 \cite{lss}. It measures not only the angular diameter distance, $D_A(z)$, but also the expansion rate of the universe, $H(z)$, which is powerful for studying dark energy \cite{task}. Since the current BAO data are not accurate enough for extracting the information of $D_A(z)$ and $H(z)$ separately \cite{okumura}, one can only determine an effective distance \cite{baosdss}: $D_v(z)=[(1+z)^2D_A^2(z)cz/H(z)]^{1/3}$. In this paper we use the BAO measurement given by the SDSS-II survey \cite{bao-SDSSII}.

In our analysis, we add a Gaussian prior on the current Hubble constant given by ref. \cite{h0};
$H_0 = 74.2 \pm 3.6$ km\,s${}^{-1}$\,Mpc${}^{-1}$ (68\% C.L.). The quoted error includes both statistical and
systematic errors. This measurement of $H_0$ is obtained from the magnitude-redshift relation of 240 low-z Type Ia supernovae at $z < 0.1$ by the Near Infrared Camera and Multi-Object Spectrometer (NICMOS) Camera 2 of the Hubble Space Telescope (HST). This is a significant improvement over the previous prior, $H_0 = 72 \pm 8$ km\,s${}^{-1}$\,Mpc${}^{-1}$, which is from the Hubble Key project final result. In addition, we impose a weak top-hat prior on the Hubble parameter: $H_0 \in [40, 100]$ km\,s${}^{-1}$\,Mpc${}^{-1}$.

Finally, we include data from Type Ia supernovae, which consists of luminosity distance measurements as a function of redshift, $D_L(z)$. In this paper we use the latest SN data sets from the Supernova Cosmology Project, ``Union Compilation 2.1'', which consists of 580 samples and spans the redshift range $0 \leq z\leq1.55$ \cite{union2.1}. This data set also provides the covariance matrix of data with and without systematic errors. In order to be conservative, we use the covariance matrix with systematic errors. When calculating the likelihood from SN, we marginalize over the absolute magnitude M, which is a nuisance parameter, as done in refs. \cite{SNMethod}.

\subsubsection{Future Measurements}

Since the current observations can not give the conclusive conclusion, we also use the simulated data from future observations of Planck and Euclid, as well as the SN sample. %Our assumptions about the future observations are close to those that could be achieved from a next generation survey such as Euclid, but do not exactly match the default index of Euclid mission.

For the CMB simulation we consider a simple full-sky ($f_{\rm sky} = 1$) simulation at Planck-like sensitivity. We neglect foregrounds and assume the isotropic noise with variance $N^{\rm TT}_\ell = N^{\rm EE}_\ell / 2 = N^{\rm BB}_\ell / 2 = 3\times10^{-4}\mu K^2$ and a symmetric Gaussian beam of 7 arcminutes full-width half-maximum (FWHM). We use the simulated $C_\ell$ up to $\ell = 2500$ for temperature \footnote{ We also use a more conservative choice $\ell_{\rm max}=1500$ to simulate the mock temperature power spectrum and find that the obtained constraints on cosmological parameters are almost unchanged.} and $\ell = 1500$ for polarization.

For the future LSS survey, we simply consider the Euclid-like survey which will measure $\sim 10^8$ galaxies with the redshifts $z< 2$. In the measurements of large scale matter power spectrum of galaxies there are generally two statistical errors: sample variance and shot noise. The uncertainty due to statistical effects, averaged over a radial bin $\Delta k$ in Fourier space, is \cite{lsserror}:
\begin{equation}
\left(\frac{\sigma_P}{P}\right)^2=2\times\frac{(2\pi)^3}{V_{\rm survey}}\times\frac{1}{4\pi k^2\Delta k}\left(1+\frac{1}{\bar{n}P}\right)^2~.
\end{equation}
The initial factor of 2 is due to the real property of the density field, $V_{\rm survey}$ is the survey volume,
and $\bar{n}$ is the mean galaxy density. In our simulations for simplicity and to be conservative,
we use only the linear matter power spectrum up to $k_{\rm max} = 0.1h$\,Mpc${}^{-1}$.
%For simplification, here we do not consider the effect of galaxy bias and neglect the correlations among bins.

For the future supernovae observation, we simulate 4000 SN samples distributed in 17 bins from $z=0.1$ to $z=1.7$. We also include $300$ low-z SN from the Nearby Supernova Factory \cite{WoodVasey:2004pj} to improve the constraints. For all the supernovae samples, we assume the variance $\sigma=0.15$ in each redshift bin.

%Results===============================================================

\section{Numerical Results}\label{results}
\subsection{Current Data}

In this section, we present the numerical results from current observational data sets
with different input theoretical models. Here, we mainly focus on the EoS of dark energy,
the primordial power spectrum index, and the total neutrino mass. %By comparing the
%constraints from different runs, we present the artificial shift in the constraints due to the theoretical correlation between cosmological parameters.

%\iffalse
%The constraints of cosmological parameters given by global fitting include two components: the mean values and the variances. In order to quantitate the difference of the results, we adopt the mean squared error ($MSE$) for comparison. For cosmological parameter, $MSE$ is the sum of the squared change of mean values and it's variance given by different input theoretical models. Taking $x$ as an example, \be MSE(x)\equiv \sqrt{(x_{i}-x_0)^2 + (\sigma_{x_{i}} - \sigma_{x_0} )^2},\ee where $x_{0}$ and $x_{i}$ are the mean values for a specific cosmological parameter given by fitting with different input models, and $\sigma_{x_{0}}$ and $\sigma_{x_{i}}$ are the corresponding variance. By comparing the $MSE$ value for different runs, we can estimate the influence from the correlation. Also, from $MSE$ value, we can classify the artificial shift of the constraints, for example, if the $MSE$ value is the mean value shift dominant, it means that the correlation is important for the measurement, since the constraints will be biased heavily without including such correlation. While if the $MSE$ value is the variance shift dominant, it means that the effects from the correlations are mainly introduce the uncertainty.\fi

\subsubsection{Dark Energy}\label{sec-DE}
\begin{table}%\hspace{-5mm}
TABLE I. $1\,\sigma$ constraints on some cosmological parameters from Union2.1+WMAP7+BAO+HST with different input theoretical models.
\begin{center}%\hspace{-15mm}

\begin{tabular}{c|c|c|c}

\hline\hline

%models& \multicolumn{2}{|c|}{$\Lambda$CDM} &\multicolumn{2}{|c} {dynamical de models} \\
models &$w_0$&$w_a$ &$\Delta\chi^2$\\
\hline
$CPL$&$-0.94\pm 0.18$&$-0.40\pm 0.78$&$-$ \\
\hline
$CPL\,\oplus\,\Omega_k$&$-0.93\pm 0.19$&$-0.47^{+1.00}_{-0.97}$&$-0.2$ \\
\hline
$CPL\,\oplus\,m_{\nu}$&$-0.92\pm 0.20$&$-0.86\pm 1.00$&$\sim 0$ \\
\hline
$CPL\,\oplus\,r$&$-0.97\pm 0.18$&$-0.11^{+0.76}_{-0.75}$&$\sim 0$\\
\hline\hline
\end{tabular}
\end{center}
\end{table}

To study the correlations between the EoS of dark energy and other cosmological parameters, such as the curvature of universe, the total neutrino mass, and tensor perturbation mode, we perform several global analyses with different input cosmological models, for example, the flat universe with massless neutrino (noted as ``$CPL$''), the non-flat universe with massless neutrino (noted as ``$CPL\,\oplus\,\Omega_k$''), the flat universe with massive neutrino model (noted as ``$CPL \,\oplus\, m_{\nu}$'') and the flat universe with tensor perturbation (noted as ``$CPL \oplus r$''). We summarize the constraints on $w_0$ and $w_a$ in table I.

\begin{figure}[t]
\begin{center}
\includegraphics[scale=0.6]{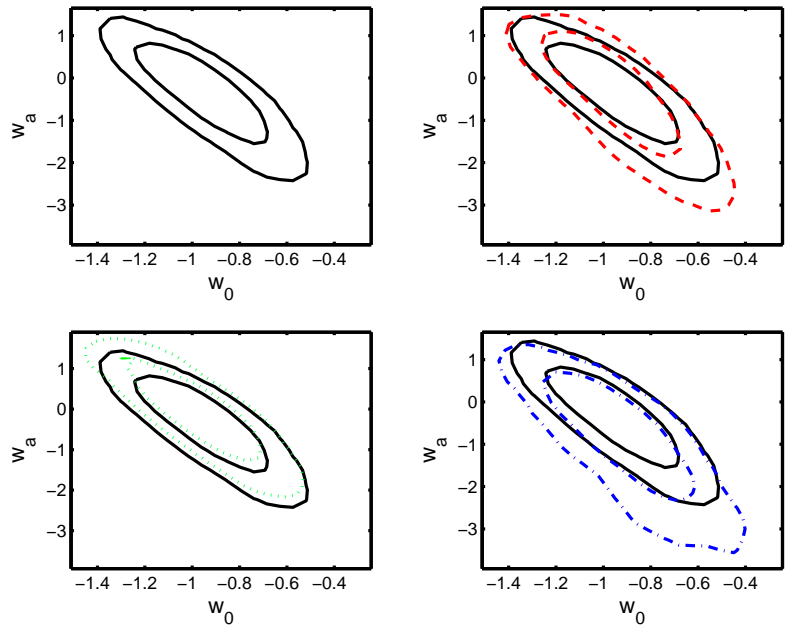}
\caption{2-dimensional cross correlation between $w_0$ and $w_a$ in different input cosmological models: the black solid lines are given by the standard $CPL$ model, the red dashed, blue dash-dot and green dotted lines are given by $CPL\,\oplus\,\Omega_k$, $CPL\,\oplus\,M_{\nu}$, and $CPL\,\oplus\,r$ models, respectively.\label{w0wa}}
\end{center}
\end{figure}

In figure \ref{w0wa}, we plot the 2-dimensional constraints of $w_0$ and $w_a$
with different input cosmological models. The upper left plot is obtained from
the standard ``$CPL$'' model. The 68\% C.L. constraints are $w_0=-0.94\pm 0.18$
and $w_a=-0.40\pm 0.78$, which is consistent with previous works \cite{wmap7}.
However, if we choose a different input cosmological model, the constraints will
change obviously. In the upper right panel of figure \ref{w0wa}, we compare the
constraints on $w_0$ and $w_a$ from the flat (black solid lines) and non-flat
(red dashed lines) universe cases, respectively. Apparently, the model with
non-zero $\Omega_k$ give weaker constraints on dark energy EoS parameters and
the median values are shifted, namely, the constraints are $w_0=-0.93\pm 0.19$
and $w_a=-0.47^{+1.00}_{-0.97}$ at the 68\% confidence level. We know that the dark energy
and the curvature are correlated via the distance relation. The effect on the distance
of dark energy EoS can be mimicked by the non-zero $\Omega_k$. Therefore, the degeneracy
between them should be very strong. In our analysis, we find that the correlation
coefficients are $-0.32$ and $0.58$ between $\Omega_k$ and $w_0$ or $w_a$, respectively. %Unfortunately, the current observations are not accurate enough. Due to the large error bars of $w_0$ and $w_a$, the non-zero curvature does not affect the constraints on the EoS of dark energy seriously.

Then, we consider the effect of tensor perturbation mode on the constraints of dark energy EoS.
In the lower left panel of figure \ref{w0wa}, we present the constraint on the EoS parameters
of dark energy from the models including tensor perturbation mode (green dotted lines).
The tensor perturbation mode mainly contributes the CMB temperature anisotropies on the
very large scales. Meanwhile, the dark energy component also affects the low multipoles
temperature power spectrum of CMB through the late-time integrated Sachs-Wolfe (ISW)
effect \cite{isw}. When including the tensor perturbation mode, we obtain the $1\,\sigma$
constraints: $w_0=-0.97\pm 0.18$ and $w_a=-0.11^{+0.76}_{-0.75}$. The constraints do not change
significantly, considering the large error bars. %The correlation coefficients are $-0.16$ and $0.27$ between $r$ and $w_0$ or $w_a$, respectively, which means the degeneracy is not very strong.

Finally, we investigate the degeneracy between the EoS of dark energy and the massive neutrino. In the lower right panel of figure \ref{w0wa}, we show the constraint on $w_0$ and $w_a$ when considering the massive neutrinos (blue dash-dot lines). In this case, the constraints on the EoS parameters of dark energy are obviously weakened and the median values are shifted: $w_0=-0.92\pm 0.20$ and $w_a=-0.86\pm1.00$ (68\% C.L.). This results in a strong anti-correlation between $\sum{m_\nu}$ and the EoS parameters of dark energy, which mainly comes from the geometric feature of our universe \cite{Hannestad:2005gj}. In addition, dynamical dark energy will modify the time evolving potential wells which affect CMB power spectra through the late time ISW effect. Dynamical dark energy can leave imprints on CMB, LSS power spectra, and Hubble diagram, nonetheless these features can be mimicked by cosmic neutrino to some extent \cite{copeland}. In our analysis, we find that the correlation coefficients are $0.08$ and $-0.37$ between $\sum{m_\nu}$ and $w_0$ or $w_a$, respectively.

\subsubsection{Inflationary parameters}
\label{sec_IF}

Currently, the cosmological observational data are in good agreement with a Gaussian, adiabatic, and scale-invariant primordial spectrum, which are consistent with single-field slow-roll inflation predictions. Measuring the spectral index $n_s$ of primordial power spectrum is important for the studies of inflation paradigm. In the literature, the constraint on $n_s$ is usually obtained by fitting the basic 6-parameter $\Lambda$CDM model. However, the constraints can be changed by the correlations between $n_s$ and other cosmological parameters, such as the running of spectral index $\alpha_s$, defined as $\alpha_s=d n_s/d\ln k$, the tensor perturbation mode, as well as the EoS parameters of dark energy.

To investigate these correlations, we constrain $n_s$ in several different input cosmological models: the standard $\Lambda$CDM model which assumes a pure scale invariant power-law power spectrum (noted as $\Lambda$CDM), the standard $\Lambda$CDM model with running spectral index (noted as ``$\Lambda$CDM$\,\oplus\,\alpha_s$''), the standard $\Lambda$CDM model with tensor perturbation mode (noted as ``$\Lambda$CDM$\,\oplus\,r$''), the standard $\Lambda$CDM model with running spectral index and tensor perturbation mode (noted as ``$\Lambda$CDM$\,\oplus\,\alpha_s\,\oplus\,r$''),  the standard $\Lambda$CDM model with running spectral index and massive neutrinos (noted as ``$\Lambda$CDM$\,\oplus\,\alpha_s\,\oplus\,m_{\nu}$''), the standard $\Lambda$CDM model with tensor perturbation mode and massive neutrinos (noted as ``$\Lambda$CDM$\,\oplus\,r\,\oplus\,m_{\nu}$''). Furthermore, we have studied the correlations between the dynamical dark energy and inflationary parameters (noted as ``$CPL$'', ``$CPL\,\oplus\,\alpha_s$'' and ``$CPL\,\oplus\,r$'' models, respectively). We list the numerical results in table II and plot the 1-dimensional distributions of $n_s$ when using different input cosmological models.

\begin{table}%\hspace{-5mm}
TABLE II. $1\,\sigma$ constraints on the Inflationary parameters $n_s$, $\alpha_s$,
and $r$ from Union2.1+WMAP7+BAO+HST. For the weakly constrained parameters, we quote
the $95\%$ upper limits instead.
\begin{center}%\hspace{-15mm}

\begin{tabular}{c|c|c|c|c}

\hline\hline

&$n_s$ &$\alpha_s$&$r$&$\Delta\chi^2$ \\
\hline
$\Lambda$CDM&$0.969\pm0.011$&$-$&$-$&$-$\\
\hline
$\Lambda$CDM$\,\oplus\,\alpha_s$&$0.951\pm0.020$&$-0.018\pm0.016$&$-$&$-2.0$ \\

\hline
$\Lambda$CDM$\,\oplus\,r$&$0.974\pm0.012$&$-$&$<0.15$&$-0.5$ \\

\hline
$\Lambda$CDM$\,\oplus\,\alpha_s\,\oplus\,r$&$0.946^{+0.020}_{-0.021}$&$-0.038\pm0.022$&$<0.37$&$-2.2$ \\

\hline
$\Lambda$CDM$\,\oplus\,m_{\nu}\,\oplus\,\alpha_s$&$0.960^{+0.021}_{-0.022}$&$-0.012\pm0.017$&$-$&$-1.6$ \\

\hline
$\Lambda$CDM$\,\oplus\,m_{\nu}\,\oplus\,r$&$0.960^{+0.024}_{-0.023}$&$-$&$<0.39$&$-1.6$ \\

\hline
$CPL$&$0.966\pm0.013$&$-$&$-$&$-0.3$ \\

\hline
$CPL\,\oplus\,\alpha_s$&$0.930^{+0.025}_{-0.026}$&$-0.029^{+0.019}_{-0.020}$&$-$&$-2.7$ \\

\hline
$CPL\,\oplus\,r$&$0.979^{+0.016}_{-0.018}$&$-$&$<0.20$&$-0.2$ \\

\hline\hline
\end{tabular}
\end{center}
\end{table}

\begin{figure}[t]
\begin{center}
\includegraphics[scale=0.5]{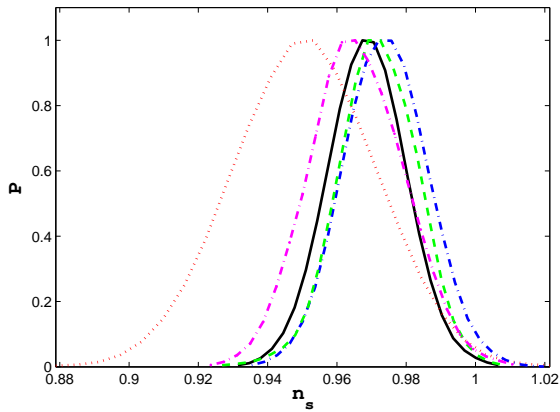}
\caption{1-dimensional probability distribution of the spectral index $n_s$ in different input cosmological models. The black solid line is given by the standard $\Lambda$CDM model, while the red dotted, blue dash-dotted line, green dashed are obtained from the $\Lambda$CDM$\,\oplus\,\alpha_s$, $\Lambda$CDM$\,\oplus\,r$ and $\Lambda$CDM$\,\oplus\,m_{\nu}$ models, respectively. For comparison, we also show the distribution from the standard $CPL$ model (purple dash-dot-dot line).\label{ns}}
\end{center}
\end{figure}

\begin{figure}[t]
\begin{center}
\includegraphics[scale=0.6]{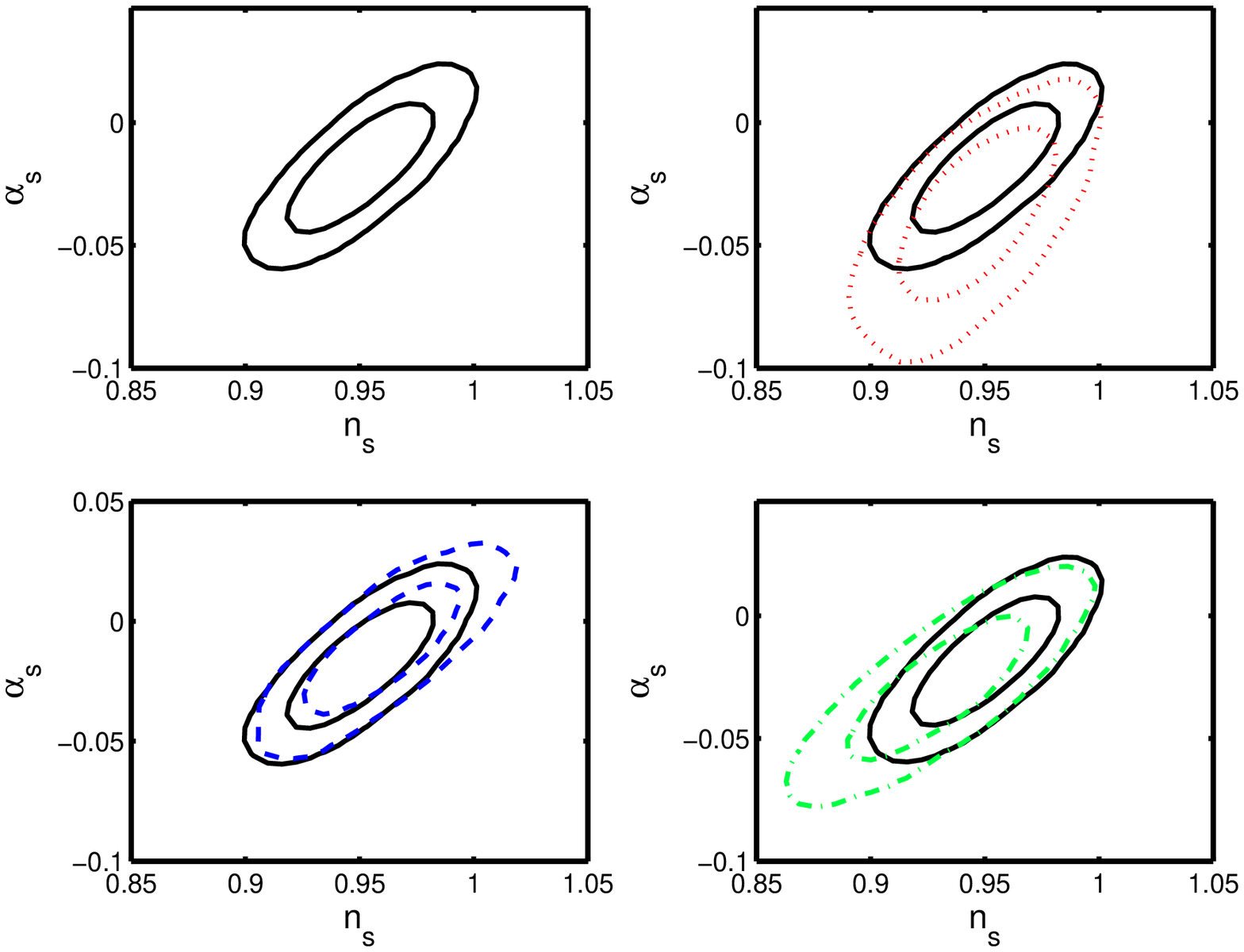}
\caption{2-dimensional cross correlation between $n_s$ and $\alpha_s$ in different input cosmological models: The black solid lines are given by the standard $\Lambda$CDM model, the red dotted lines, blue dash-dot lines and green dashed lines are given by $\Lambda$CDM$\,\oplus\,\alpha_s\,\oplus\,r$, $\Lambda$CDM$\,\oplus\,m_{\nu}\,\oplus\,\alpha_s$ and $CPL\,\oplus\,\alpha_s$ models, respectively.\label{ns-as}}
\end{center}
\end{figure}

In the standard 6-parameter $\Lambda$CDM model, we obtain the $1\,\sigma$ constraint on $n_s$ is $n_s=0.969\pm0.011$, which is consistent with previous works \cite{wmap7}. When considering the running of spectral index, the constraint of $n_s$ is significantly weakened, $n_s=0.951\pm0.020$ (68\% C.L.), while the 68\% constraint of $\alpha_s$ is $\alpha_s=-0.018\pm0.016$. The change of the constraint on $n_s$ is about $1.6\,\sigma$. We also obtain that the correlation coefficient between $n_s$ and $\alpha_s$ is $0.83$, which implies the strong correlation between them. When including tensor perturbation mode, the constraint on $n_s$ slightly changes: $n_s=0.974\pm0.012$ at the 68\% confidence level. When varying both $\alpha_s$ and $r$ simultaneously, the $1\,\sigma$ error bars of Inflationary parameters are significantly enlarged: $n_s=0.946^{+0.020}_{-0.021}$ and $\alpha_s=-0.038\pm0.022$. The correlation coefficients among $n_s$, $\alpha_s$ and $r$ are $0.70$, $0.35$ and $-0.60$, respectively. In figure \ref{ns-as} we plot the 2-dimensional constraints on the panel ($n_s$,$\alpha_s$) in the upper two panels with or without considering the tensor perturbation, respectively.

The correlation between the early universe inflation and late time accelerating expansion can also affect the constraints on inflationary parameters. In time-evolving dark energy model ($CPL$), we obtain the 68\% C.L. constraint $n_s=0.966\pm0.013$, which is similar with that in the $\Lambda$CDM framework. However, when considering the running of spectral index, the constraints of $n_s$ and $\alpha_s$ are obviously weakened, namely, $n_s=0.930^{+0.025}_{-0.026}$, and $\alpha=-0.029^{+0.019}_{-0.020}$ at the 68\% confidence level. More importantly, their mean values are shifted significantly, when comparing them from the standard $\Lambda$CDM model. In the ``$CPL\,\oplus\,r$'' model, we obtain the constraints $n_s=0.979^{+0.016}_{-0.018}$  ($1\,\sigma$) and $r<0.20$ ($2\,\sigma$). Their error bars become larger which is clearly shown in the lower right panel of figure \ref{ns-as}, due to the correlation between the dark energy and the tensor perturbation we discuss before.

We also study the correlations between the massive neutrino and $n_s$, $\alpha_s$, $r$. With the massive neutrino, we obtain the constraints $n_s=0.960^{+0.024}_{-0.023}$  ($1\,\sigma$) and $r<0.39$ ($2\,\sigma$) in the ``$\Lambda$CDM$\,\oplus\,m_{\nu}\,\oplus\,r$'' model, whose error bars are much larger than those in the massless neutrino case.

\subsubsection{Neutrino mass}
\label{sec-NM}

\begin{table}%\hspace{-5mm}
TABLE III. The 95\% confidence level upper limits on the total mass of neutrinos from Union2.1+WMAP7+BAO+HST.
\begin{center}%\hspace{-15mm}

\begin{tabular}{c|c|c}

\hline\hline

&$\Sigma m_{\nu}$&$\Delta\chi^2$ \\

\hline
$LCDM\oplus m_{\nu}$&$<0.45\,$eV &$-$\\
\hline
$LCDM\oplus m_{\nu}\oplus r$&$<0.55\,$eV &$-1.1$\\
\hline
$LCDM\oplus m_{\nu}\oplus \alpha_s$&$<0.42\,$eV &$-1.1$\\

\hline
$CPL\oplus m_{\nu}$&$<0.81\,$eV &$\sim 0$\\

\hline\hline
\end{tabular}
\end{center}
\end{table}

\begin{figure}[t]
\begin{center}
\includegraphics[scale=0.5]{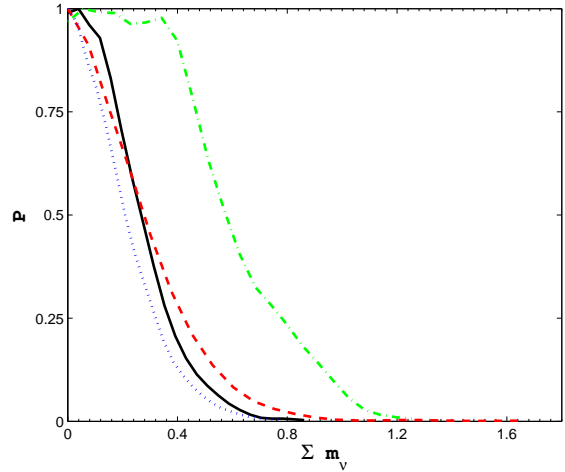}
\caption{1-dimensional probability distributions of the neutrino mass in different input cosmological models. The black solid line is given by the standard $\Lambda$CDM model, while the blue dotted, red dashed and the green dash-dot lines are given by the $\Lambda$CDM$\,\oplus\,m_{\nu}\,\oplus\,\alpha_s$, $\Lambda$CDM$\,\oplus\,m_{\nu}\,\oplus\,r$ and $CPL\,\oplus\,m_{\nu}$ models, respectively.\label{mu}}
\end{center}
\end{figure}

Neutrino, as a hot dark matter candidate, can be constrained from the free streaming modification of the transfer function of the matter power spectrum. Thus, the constraints on neutrino mass are correlated with the shape of primordial power spectrum. Since the free streaming effects can be compensated by the modification of the primordial spectrum, the constraints on total neutrino mass can be influenced by the inflationary parameters, as we discuss above. Moreover, the behaviour of dark energy EoS can also affect the constraint of the total neutrino mass. In table III and figure \ref{mu} we show the numerical constraints on the massive neutrino from different input models: the standard $\Lambda CDM$ model (noted as ``$\Lambda$CDM$\,\oplus\,m_{\nu}$''), the standard $\Lambda$CDM with tensor perturbation mode (noted as ``$\Lambda$CDM$\,\oplus\,m_{\nu}\,\oplus\,r$''), the standard $\Lambda$CDM with running of spectral index (noted as "$\Lambda$CDM$\,\oplus\,m_{\nu}\,\oplus\,\alpha_s$''), and the $CPL$ dynamical dark energy model (noted as ``$CPL\,\oplus\,m_{\nu}$'').

In the framework of standard $\Lambda$CDM model, the $2\,\sigma$ upper limit on the total neutrino mass is $\sum m_{\nu} < 0.45\,$eV. When including the tensor perturbation mode, the constraint is slightly weakened, namely, $\sum m_{\nu} < 0.55\,$eV at the 95\% confidence level. As we know, both the negative $\alpha_s$ and the massive neutrino lead to a damped power on small scales. The effect of massive neutrinos may be compensated by a non-vanishing running of the primordial spectrum. On the other hand, the running is negative, this will lead to even more stringent constraints on the neutrino mass compared with fittings in the constant scalar spectral index cosmology \cite{fnurun}. In practice, we find the 95\% upper limit on the total neutrino mass becomes smaller, when varying the running of the spectral index, $\sum m_{\nu} < 0.42\,$eV.

The massive neutrino is strongly correlated with the EoS of dark energy. Therefore, the behaviour of dark energy model, especially the dynamical dark energy model, will significantly enlarge the constraint of total neutrino mass. In the ``$CPL\,\oplus\,m_{\nu}$'' model, we obtain the $2\,\sigma$ upper limit on the massive neutrino is $\sum m_{\nu} < 0.81\,$eV, as shown in figure \ref{mu}.

%Fitting with simulated future data
\subsection{Future Data}

From the results presented above, we see that there are some strong degeneracies among cosmological parameters, such as $w$ and $\sum{m_\nu}$, $n_s$ and $\alpha_s$. Due to the precision of current observations, these degeneracies can not be broken and will weaken the constraints of cosmological parameters. Therefore, it is worthwhile discussing whether future observations could give more stringent constraints on the cosmological parameters and break these degeneracies efficiently. For this purpose, we have performed a further analysis and we have chosen the fiducial model in perfect agreement with current data: $\Omega_bh^2=0.0227$, $\Omega_{c}h^2=0.114$, $\vartheta=1.041$, $\tau=0.0865$, $w_0=-1$, $w_a=0$, $n_s=1$, $A_s=2.85\times 10^{-9}$ at $k_\ast=0.05\,$Mpc${}^{-1}$.

\subsubsection{$\Omega_k$ and $w(z)$}

\begin{table}%\hspace{-5mm}
TABLE IV. The $1\,\sigma$ constraints on parameters from the simulated mock data with the fiducial value $\Omega_k=-0.013$.
\begin{center}%\hspace{-15mm}

\begin{tabular}{c|c|c|c|c}

\hline\hline

%models& \multicolumn{2}{|c|}{$\Lambda$CDM} &\multicolumn{2}{|c} {dynamical de models} \\
models &$w_0$&$w_a$&$\Omega_K$&$\Delta\chi^2$\\
\hline
fiducial model&$-1$&$0$&$-0.013$&$-$\\
\hline
$CPL\,\oplus\, \Omega_k$&$-1.000^{+0.0270}_{-0.0280}$&$0.00\pm0.130$&$-0.0130^{+0.0017}_{-0.0020}$&$0$ \\
\hline
$CPL$&$-1.068^{+0.0210}_{-0.0220}$&$0.481^{+0.074}_{-0.072}$& set to 0&$23$ \\
\hline
Constant $w\,\oplus\, \Omega_k$&$-1.0000\pm0.0085$&set to 0&$-0.0127^{+0.0011}_{-0.0010}$&$0$ \\
\hline
Constant $w$&$-0.9407^{+0.0057}_{-0.0058}$&set to 0& set to 0&$41$ \\
\hline\hline
\end{tabular}
\end{center}
\end{table}

We use the fiducial mock data to study the degeneracy between the EoS of dark energy and $\Omega_k$. When we simulate the future data, we choose the fiducial value of curvature $\Omega_k=-0.013$ which is allowed by current observational data. We summarize the numerical results in table IV.

\begin{figure}[t]
\begin{center}
\includegraphics[scale=0.6]{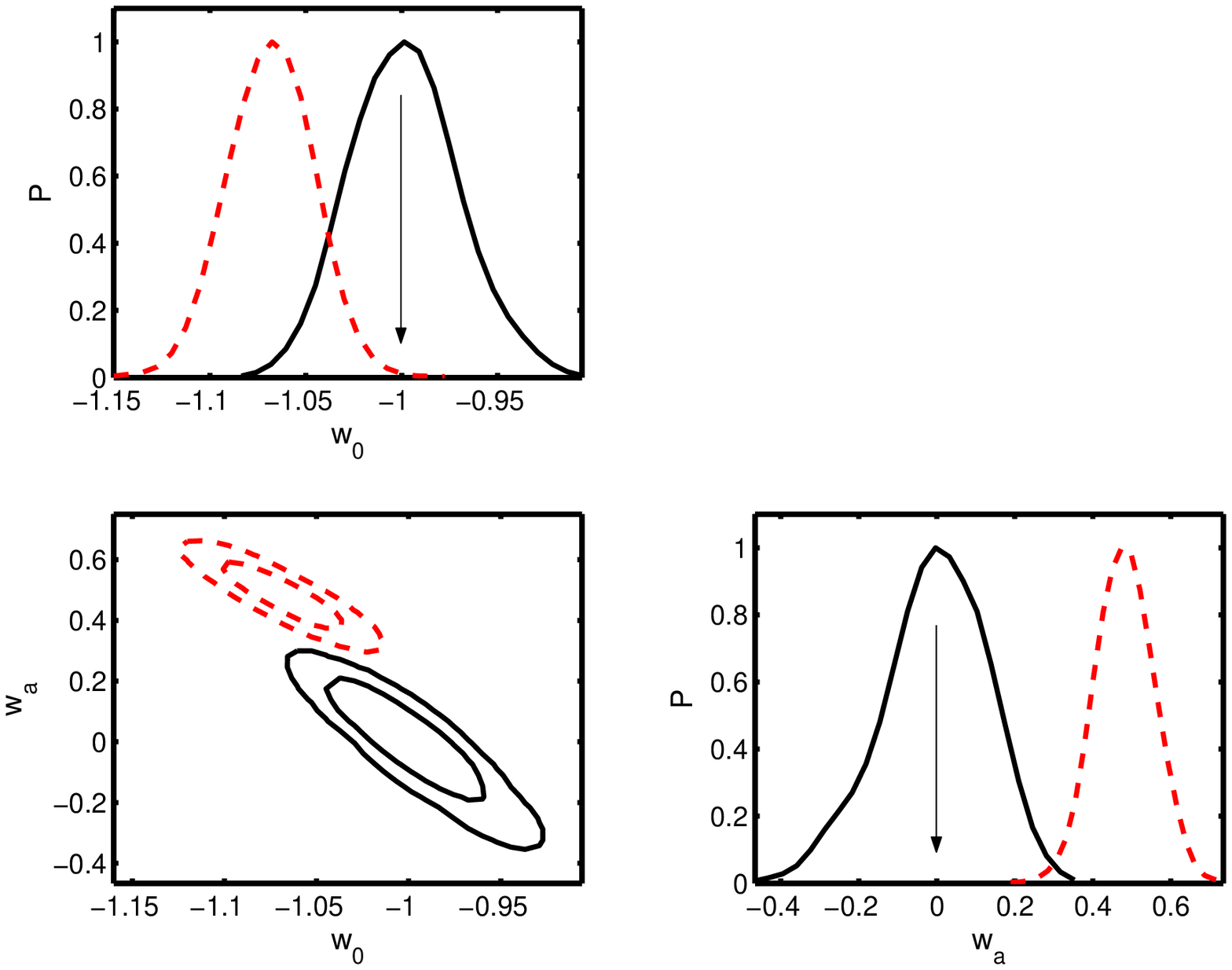}
\caption{1 and 2-dimensional constraints on the panel ($w_0$,$w_a$) from the simulated mock data with the fiducial value $\Omega_k=-0.013$. The black solid lines are obtained by using the correct model which allows $\Omega_{k}$ varying. For comparison, we also use a biased model which fixes the curvature $\Omega_k\equiv0$ to constrain the EoS of dark energy, shown as the red dashed lines.\label{fc_w0wa_omkd13}}
\end{center}
\end{figure}

Firstly, we use the correct input cosmological model ``$CPL\,\oplus\,\Omega_k$'' to constrain parameters. By combining the future mock data from PLANCK, Euclid and supernovae, we obtain the constraints on the EoS parameters: $w_0=-1.000^{+0.0270}_{-0.0280}$ and $w_a=0.00\pm0.130$ at the 68\% confidence level. When comparing the constraints from the current observations, the future data constrain the EoS parameters much more stringently. The error bars of $w_0$ and $w_a$ are shrunk by a factor of 6. In figure \ref{fc_w0wa_omkd13} we plot the 2-dimensional constraint on the panel ($w_0$,$w_a$) (black solid lines). Meanwhile, the mock data also give very tight constraint on the curvature: $\Omega_k = -0.0130^{+0.0017}_{-0.0020}$ (68\% C.L.).

%We then do the fitting with the forecast data sets. Fitting with the true model, which takes $\Omega_K$ as a free parameter, we can estimate the variance on $w_0$ and $w_a$, and fitting with a biased model, which we fix $\Omega_K=0$, we will get a biased constraints.

Besides the correct input model, we also use a biased model to fit the mock data, in order to study the degeneracy between $\Omega_k$ and $w$. Here, we assume the flat universe ($\Omega_k\equiv0$). Due to the strong degeneracy between them, in this case the constraints of $w_0$ and $w_a$ are quite different from those of the correct input model. The $1\,\sigma$ constraints are $w_0=-1.068^{+0.0210}_{-0.0220}$ and $w_a=0.481^{+0.074}_{-0.072}$. One can see that the error bars are smaller than above, due to less free parameters in this case. More importantly, the mean values of $w_0$ and $w_a$ are shifted obviously, as shown in the red dashed lines of figure \ref{fc_w0wa_omkd13}. When considering the error bars, these mean values will be ruled out at more than $3\,\sigma$ confidence level, which implies that the future observations could break this degeneracy very well.

\begin{figure}[t]
\begin{center}
\includegraphics[width=120mm,height=60mm]{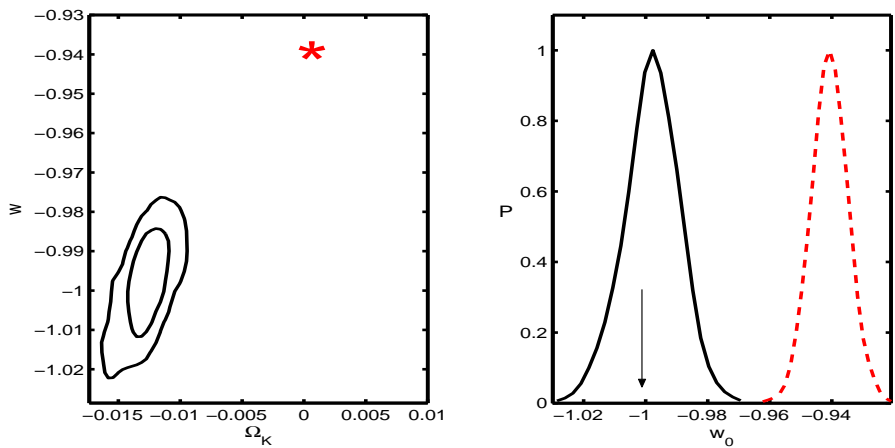}
\caption{1-dimensional constraint on $w$ and 2-dimensional constraint on the panel ($\Omega_k$,$w$) from the simulated mock data with the fiducial value $\Omega_k=-0.013$. The black solid lines are obtained by using the correct model which allows $\Omega_{k}$ varying, while the red star denotes the best fit model when assuming the flat universe.\label{fc_womk_omkd13}}
\end{center}
\end{figure}

We also use the constant $w$ model ($w_a\equiv0$) to fit the mock data again and plot the results in figure \ref{fc_womk_omkd13}. When varying $\Omega_k$, we obtain the constraints: $w=-1.0000\pm0.0085$ and $\Omega_k=-0.0127^{+0.0011}_{-0.0010}$ (68\% C.L.), which recover the fiducial model very well. However, when assuming the flat universe, due to the degeneracy, the $1\,\sigma$ constraint on $w$ is shifted: $w=-0.9407^{+0.0057}_{-0.0058}$, which will be excluded by the future data at more than $6\,\sigma$ confidence level. In this case, the minimal $\chi^2$ are very large, which implies that this kind of model does not fit the mock data well.

%With the mocked data, we get $w_0= -0.939\pm 0.036$ and $w_a=-0.404\pm 0.143$ for fitting with the fiducial model, and $w_0= -0.942\pm 0.042$ and $w_a=-0.390\pm 0.208$ for including $\Omega_K$ as free parameter additionally. We find that the correlation between EOS of dark energy and $\Omega_K$ can not influence the constraints on $w_0$ and $w_a$ seriously, and the main effects are enlarging the variance of EoS of dark energy.

\subsubsection{$\sum m_{\nu}$ and $w$}

\begin{table}%\hspace{-5mm}
TABLE V. The $1\,\sigma$ constraints on parameters from the simulated mock data with the fiducial value $\sum m_{\nu}=0.4\,$eV.
\begin{center}%\hspace{-15mm}

\begin{tabular}{c|c|c|c|c}

\hline\hline

%models& \multicolumn{2}{|c|}{$\Lambda$CDM} &\multicolumn{2}{|c} {dynamical de models} \\
models &$w_0$&$w_a$&$\sum m_{\nu}$&$\Delta\chi^2$\\
\hline
fiducial model&$-1$&$0$&$0.4\,$eV&$-$\\
\hline
$CPL\,\oplus\, m_{\nu}$&$-1.000\pm 0.0240$&$0.00\pm 0.110$&$0.400^{+0.020}_{-0.017}$&$0$ \\
\hline
$CPL$&$-1.031\pm0.0140$&$0.589\pm 0.043$&set to 0&$40$ \\
\hline
Constant $w\,\oplus\, m_{\nu}$&$-1.001^{+0.0130}_{-0.0120}$& set to 0&$0.400^{+0.011}_{-0.015}$&$0$ \\
\hline
Constant $w$&$-0.8503^{+0.0037}_{-0.0038}$&set to 0& set to 0&$104$ \\
\hline\hline
\end{tabular}
\end{center}
\end{table}
%The correlations between neutrino mass and EOS can not be negligible for constraining EOS. In massive neutrino models, we get $-0.929\pm 0.051$ and $w_a=-0.528\pm 0.290$, comparing with those given by fitting with the fiducial model, $w_0=-0.939\pm 0.036$ and $w_a=-0.404\pm 0.143$, we find that both the mean values and the variance of $w_0$ and $w_a$ are shifted seriously, especially, for $w_a$, the mean values are shifted about $1\sigma$ and the variance are almost doubled. We get $MSE(w_0)=...$ and $MSE(w_a)=...$, which shows that the correlations between massive neutrino and dark energy are important for measuring the EoS of dark energy.

%To study the correlation between EoS of dark energy and neutrino mass, we chose a fiducial model with massive neutrino to simulate the data sets of future experiments, and then do the fitting for the correct model and a biased model respectively. The input cosmology is: $\Omega_b h^2=0.0227,~\Omega_{DM}h^2=0.114,~\vartheta=1.041,~\tau=0.0865,~w_0=-1,~w_a=0,~f_{\nu}=0.04,~n_s=1,~A_s=2.85\times 10^{-9}$, which is consistent with the current observations, and $f_{\nu}=0.04$ gives $\Sigma m_{\nu}\sim 0.4 ~eV$ which is within the $2\sigma$ confidence level limit from the current CMB, LSS and SN Ia data sets.

\begin{figure}[t]
\begin{center}
\includegraphics[scale=0.6]{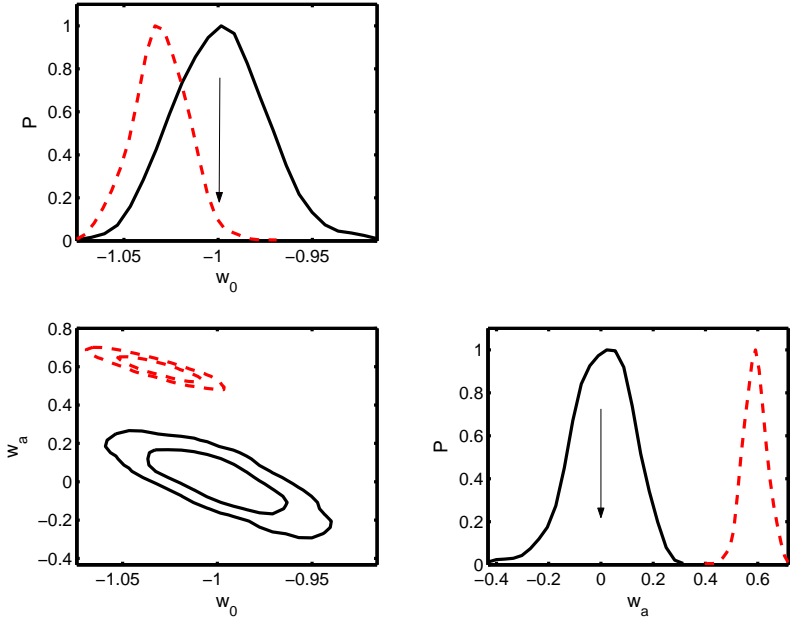}
\caption{1 and 2-dimensional constraints on the panel ($w_0$,$w_a$) from the simulated mock data with the fiducial value $\sum m_{\nu}=0.4\,$eV. The black solid lines are obtained by using the correct model which allows $\sum m_{\nu}$ varying. For comparison, we also use a biased model with the massless neutrino to constrain the EoS of dark energy, shown as the red dashed lines.\label{fs_w0wa_nud45}}
\end{center}
\end{figure}

\begin{figure}[t]
\begin{center}
\includegraphics[width=120mm,height=60mm]{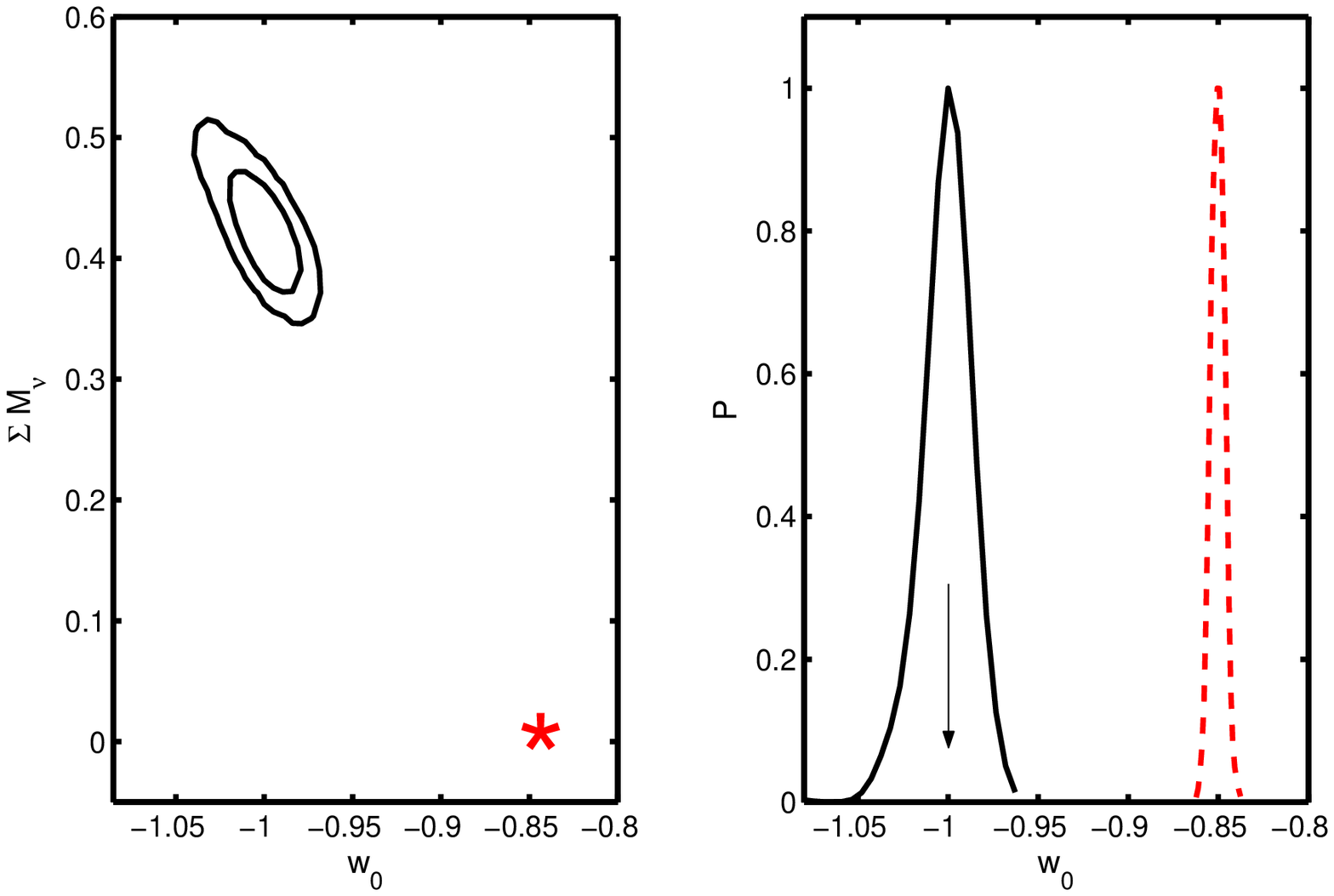}
\caption{1-dimensional constraint on $w$ and 2-dimensional constraint on the panel ($w$,$\sum m_{\nu}$) from the simulated mock data with the fiducial value $\sum m_{\nu}=0.4\,$eV. The black solid lines are obtained by using the correct model which allows $\sum m_{\nu}$ varying, while the red star denotes the best fit model when assuming the massless neutrino model.\label{fs_cw}}
\end{center}
\end{figure}

Similarly, we use the same method to study the degeneracy between $w$ and $\sum m_\nu$. When generating the simulated mock data, we choose the fiducial value of total neutrino mass $\sum m_{\nu}=0.4\,$eV, which is in agreement with the current constraint at $2\,\sigma$ confidence level.

In figure \ref{fs_w0wa_nud45}, we plot the 2-dimensional constraints on $w_0$ and $w_a$ obtained by using different input cosmological models. Firstly, we consider the correct model (``$CPL\,\oplus\, m_{\nu}$'') and obtain very tight constraints on the EoS of dark energy and the total neutrino mass: $w_0=-1.000\pm0.0240$, $w_a=0.00\pm0.110$ and $\sum m_\nu=0.400^{+0.020}_{-0.017}$eV at the 68\% confidence level. The future measurements could give conclusive conclusion about the total neutrino mass. However, if assuming the massless neutrino ($\sum m_\nu\equiv0$), we will obtain the biased constraints on the EoS of dark energy: $w_0=-1.031\pm0.0140$ and $w_a=0.589\pm0.043$ (68\% C.L.), due to the strong degeneracy between $w$ and $\sum m_\nu$. The mean values are significantly different from the input fiducial model, especially for the parameter $w_a$. Consequently, the minimal $\chi^2$ of this case will be very large, which implies that this best fit model has been ruled out by the data at more than $3\,\sigma$ confidence level. We list the numerical results in table V.

Again, we also use the constant EoS of dark energy to study this degeneracy. In the framework of the correct $w\,\oplus\,m_\nu$ model, the $1\,\sigma$ constraints are $w=-1.001^{+0.0130}_{-0.0120}$ and $\sum m_\nu=0.400^{+0.011}_{-0.015}$. When forcing $\sum m_\nu\equiv0$, the best fit model of $w$ is shifted to $w=-0.8503^{+0.0037}_{-0.0038}$, which is more than $10\,\sigma$ away from the fiducial value of $w$, as shown in figure \ref{fs_cw}. The future observations could break this degeneracy efficiently.

%We also compare the fitting results for a constant EoS, ie. we do the fitting by assuming EoS is a constant for massive and massless neutrinos respectively. We plot the cross correlation of $\Sigma M_{\nu}$ and $w$ in figure \ref{fs_cw} by the black solid lines, and we label the best fit point of the biased model which assuming massless neutrino,  $w=-0.85,~ \Sigma M_{\nu}=0$ with a red star. The unconsistancy of the two results shows the importance of the degeneracy.

\subsubsection{$n_s$, $\alpha$ and $r$}
\begin{table}%\hspace{-5mm}
TABLE VI. $1\,\sigma$ constraints on Inflationary parameters from the simulated mock data.
\begin{center}%\hspace{-15mm}

\begin{tabular}{c|c|c|c}

\hline\hline

%models& \multicolumn{2}{|c|}{$\Lambda$CDM} &\multicolumn{2}{|c} {dynamical de models} \\
models &$n_s$&$\alpha_s$&$\Delta\chi^2$ \\
\hline
fiducial model&$1$&$-0.05$&$-$ \\
\hline
correct~model&$1.0000\pm0.0010$&$-0.0502^{+0.0039}_{-0.0040}$&$0$ \\
\hline
biased~model&$1.0052^{+0.0015}_{-0.0010}$&set~to~0&$78$ \\

\hline\hline

%models& \multicolumn{2}{|c|}{$\Lambda$CDM} &\multicolumn{2}{|c} {dynamical de models} \\
models &$n_s$&$r$&$\Delta\chi^2$ \\
\hline
fiducial model&$1$&$0.2$&$-$ \\
\hline
correct~model&$1.0000^{+0.0011}_{-0.0010}$&$0.200^{+0.015}_{-0.020}$&$0$ \\
\hline
biased~model&$0.9975{\pm 0.0009}$&set~to~0&$60$ \\
\hline\hline

\end{tabular}
\end{center}
\end{table}

In order to study the degeneracies between $n_s$ and $\alpha_s$, $r$, we simulate two mock data sets: one with $\alpha_s=-0.05$ and the other one with $r=0.2$, which are in agreement with the current data. In table VI we list the constraints of these Inflationary parameters from two mock data.

\begin{figure}[t]
\begin{center}
\includegraphics[width=120mm,height=60mm]{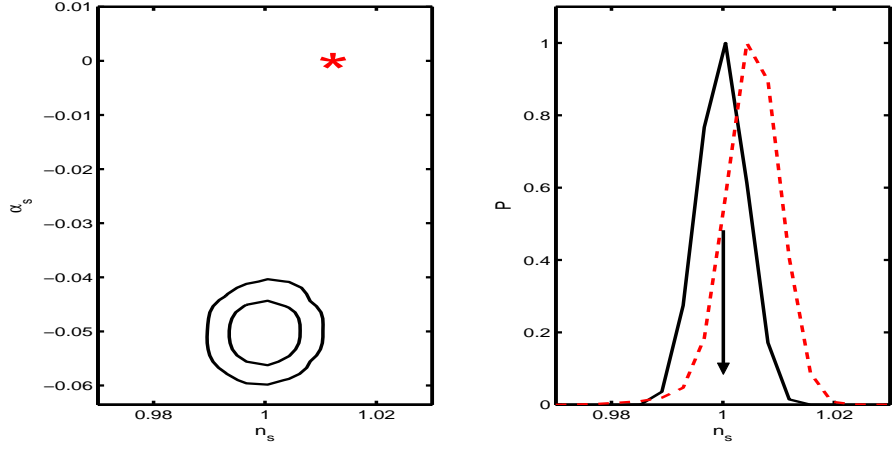}
\caption{1-dimensional constraint on $n_s$ and 2-dimensional constraint on the panel ($n_s$,$\alpha_s$) from the simulated mock data with the fiducial value $\alpha_s=-0.05$. The black solid lines are obtained by using the correct model which allows $\alpha_s$ varying, while the red star denotes the best fit model when fixing $\alpha_s=0$.\label{fc_asns}}
\end{center}
\end{figure}

Using the correct input model, we obtain very stringent constraints on $n_s$ and $\alpha_s$: $n_s=1.0000{\pm 0.0010}$, $\alpha_s =-0.0502^{+0.0039}_{-0.0040}$ (68\% C.L.), whose error bars are almost shrunk by a factor of 10. When assuming the spectral index is scale-independent ($\alpha_s\equiv0$), the constraint of $n_s$ changes: $n_s=1.0052^{+0.0015}_{-0.0010}$ at the 68\% confidence level. The mean value of $n_s$ is shifted significantly, when comparing with the obtained error bar. We compare these two results in figure \ref{fc_asns}. The obtained correlation coefficient between $n_s$ and $\alpha_s$ is $0.1$, which means this degeneracy does not show in the future data significantly.

\begin{figure}[t]
\begin{center}
\includegraphics[width=120mm,height=60mm]{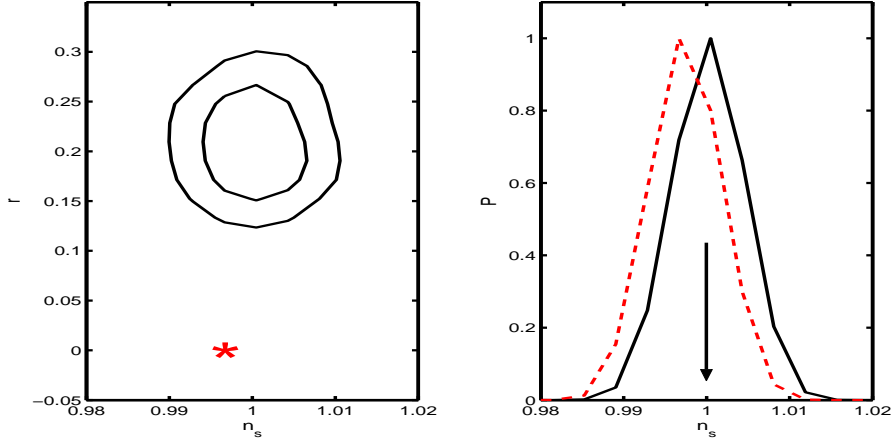}
\caption{1-dimensional constraint on $n_s$ and 2-dimensional constraint on the panel ($n_s$,$r$) from the simulated mock data with the fiducial value $r=0.2$. The black solid lines are obtained by using the correct model which allows $r$ varying, while the red star denotes the best fit model when neglecting the tensor perturbation mode.\label{fc_rns}}
\end{center}
\end{figure}

The similar situation is also found by the study about the degeneracy between $n_s$ and $r$. In figure \ref{fc_rns} we show the constraints on $n_s$ and $r$ from the correct and biased input models, respectively. When including the tensor perturbation mode, the $1\,\sigma$ constraints are: $n_s=1.0000^{+0.0011}_{-0.0010}$ and $r=0.200^{+0.015}_{-0.020}$. Neglecting the tensor perturbation, the constraint on $n_s$ is shifted to a lower value: $n_s=0.9975{\pm 0.0009}$. We obtain the correlation coefficient between $n_s$ and $r$ is $-0.05$, which implies that the degeneracies between $n_s$ and $r$ is not notable for the future data sets.

%With the data, we then do the fitting for involving/ingnoring the running respectively.  In figure \ref{fc_asns}, we plot the $1\sigma$ and $2\sigma$ cross correlation of $\alpha_s$ and $n_s$ given by the data, and $\alpha_s$ and $n_s$ are well constraints, $\alpha_s = -0.005\pm0.004,~n_s=1.00\pm0.002$, while by fitting the biased model, which do not include $\alpha_s$ as free parameter, we get $n_s=1.01\pm0.002$ and it is labeled by a red star. The $\Delta\chi^2\sim 78$ for the two models. From the shift of the mean value of $n_s$, we find that, the degeneracies between $\alpha_s$ and $n_s$ is not notable for the future data sets.

%To study the correlations between the tensor modes $r$ and $n_s$, we simulate the data sets by taking $\Omega_b h^2=0.0227,~\Omega_{DM}h^2=0.114,~\vartheta=1.041,~\tau=0.0865,~w=-1,~n_s=1,~A_s=2.2\times 10^{-9},~r=0.2$ which is comparable with the current data, and then we do the fitting for involving/ingnoring the tensor modes respectively, and see the constraints on $n_s$. In figure \ref{fc_rns} we plot the $1\sigma$ and $2\sigma$ constraints for $r$ and $n_s$ by fitting with freeing $r$, and we get the constraints are:$r=0.200\pm0.036, ~n_s=1.000\pm0.002$, however, when we fixed $r=0$ compulsively, the best fit value for $n_s$ is $n_s= 0.99$, which is labeled by the blue triangle in the plot. We compared the $\chi^2$ for the two fitting results, we find $\Delta\chi^2\simeq 60$.

\subsection{Conclusion and discussion}

In this paper we perform global fitting analyses on serval cosmological models and present the constraints on cosmological parameters from the latest astronomical observations, including WMAP7, BAO measurement from the SDSS-II survey and supernovae ``Union2.1'' sample. We find that, with different input theoretical models, the constraints on some cosmological parameters can be very different, such as the equation of state of dark energy, the total neutrino mass and the index of primordial power spectrum, due to the strong degeneracies among these parameters.

With the high quality mock data, such as future SN, CMB and LSS observations, we further study the impact of these degeneracies on the constraints of cosmological parameters. Using these accurate future data, the obtained results show that, some degeneracies can not be neglected, such as the EoS of dark energy and the curvature, the EoS and the massive neutrino. Ignoring them forcibly will lead to seriously biased constraints.

%Acknowledgments=======================================================

\section*{Acknowledgements}

We thank Xinmin Zhang and Gong-Bo Zhao for helpful discussion. We acknowledge the use of the Legacy Archive for Microwave Background Data Analysis (LAMBDA). Support for LAMBDA is provided by the NASA Office of Space Science. The calculation is taken on Deepcomp7000 of Supercomputing Center, Computer Network Information Center of Chinese Academy of Sciences. This work is supported in part by the National Science Foundation of China under Grant
Nos. 11033005, by the 973 program under Grant No. 2010CB83300, by the Chinese Academy of Science under Grant No. KJCX2-EW-W01.

%End===================================================================

\end{document}